\documentclass[aip,amsmath,amssymb,reprint]{revtex4-1}
\usepackage{epsfig}
\usepackage{graphicx, verbatim}
\usepackage{dcolumn}
\usepackage{bm}
\usepackage{epstopdf}
\usepackage[utf8]{inputenc}
\usepackage[T1]{fontenc}
\usepackage{mathptmx}
\usepackage{etoolbox}
\usepackage{hyperref}
\hypersetup{colorlinks=true,linkcolor=blue,filecolor=magenta,urlcolor=cyan}

\begin{document}

\title{Efficient characteristics of exchange coupling and spin-flop transition in Py/Gd bilayer using anisotropic magnetoresistance}

\author{Kaiyuan Zhou}
\author{Xiang Zhan}
\author{Zishuang Li}
\author{Haotian Li}
\author{Chunjie Yan}
\affiliation{Jiangsu Provincial Key Laboratory for Nanotechnology, National Laboratory of Solid State Microstructures and School of Physics, Nanjing University, Nanjing 210093, China}
\author{Lina Chen}
\thanks{Authors to whom correspondence should be addressed:\\
Lina Chen,chenlina@njupt.edu.cn; Ronghua Liu,rhliu@nju.edu.cn}
\affiliation{School of Science, Nanjing University of Posts and Telecommunications, Nanjing 210023, China}
\author{Ronghua Liu}
\thanks{Authors to whom correspondence should be addressed:\\
Lina Chen,chenlina@njupt.edu.cn; Ronghua Liu,rhliu@nju.edu.cn}
\affiliation{Jiangsu Provincial Key Laboratory for Nanotechnology, National Laboratory of Solid State Microstructures and School of Physics, Nanjing University, Nanjing 210093, China}

\begin{abstract}
The interlayer antiferromagnetic coupling rare-earth/transition-metal bilayer ferrimagnet systems have attracted much attention because they present variously unusual temperature-and field-dependent nontrivial magnetic states and dynamics. These properties and the implementation of their applications in spintronics highly depend on the significant temperature dependence of the magnetic exchange stiffness constant $A$. Here, we quantitatively determine the temperature dependence of magnetic exchange stiffness $A_{Py-Gd}$ and $A_{Gd}$ in the artificially layered ferrimagnet consisting of a Py/Gd bilayer, using a measurement of anisotropic magnetoresistance (AMR) of the bilayer thin film at different temperatures and magnetic fields. The obtained temperature dependence of $A_{Py-Gd}$ and $A_{Gd}$ exhibit a scaling power law with the magnetization of Gd. The critical field of spin-flop transition and its temperature dependence can also be directly obtained by this method. Additionally, the experimental results are well reproduced by micromagnetic simulations with the obtained parameters $A_{Py-Gd}$ and $A_{Gd}$, which further confirms the reliability of this easily accessible technique.
\end{abstract}

\maketitle

The ferromagnetic bilayers provide possibilities for designing magnetic properties significantly different from the single ferromagnet by combining two materials with contrasting characteristics~\cite{LWLi,KDobrich}. The interlayer antiferromagnetic coupling rare-earth/transition-metal (RE/TM) bilayer systems with contrasting saturation magnetizations and significantly different Curie temperatures have attracted much attention because they present variously unusual temperature- and field-dependent magnetic static states and dynamics~\cite{Sun,Paul,Ueda}. For example, the antiferromagnetic coupling between the RE and TM layers can lead to temperature and/or field-driven novel magnetic states and phase transitions between them~\cite{Higgs,Demirtas}. In addition to providing an excellent platform for fundamental magnetic research, ferrimagnetic RE-TM alloys and their multilayers exhibit substantial potential applications in spintronic with high speed, high density, and low-energy cost~\cite{Blasing,KUeda}. For example, achieving strong out-of-plane magnetic anisotropy~\cite{Radu,Ishibashi,Ding}, ultrafast laser-induced magnetization reversal~\cite{ostler,Chatterjee,Dannegger}, highly efficient spin-charge conversion and tunable magnetic damping for spin-torque driven magnetization reversal and fast motion of magnetic skyrmion or bubble~\cite{WQWang,YWang,Woo,Montoya,Lee}, and magnon chirality in these ferrimagnetic alloys and multilayers~\cite{YLiu,Blasing}. All these nontrivial magnetic states and dynamics and the implementation of their applications highly depend on the significant temperature dependence of the magnetic exchange stiffness constant $A$~\cite{Niitsu,Kuz2,Altuncevahir}.

On the other hand, micromagnetic modeling has been widely recognized as a powerfully complementary tool in many experimental measurements, especially for calculations of various spin textures, the critical field of spin-flop and spin-flip, magnetization dynamics and their spatial profiles~\cite{Chen1}. The premise for obtaining reliable simulation results consistent with the experiment is that the actual magnetic material parameters: the effective crystalline anisotropy $K$, exchange stiffness $A$, and saturation magnetization $M_S$ (the additional Gilbert damping constant $\alpha$ for spin dynamics) need to be known as input fundamental micromagnetic parameters\cite{oommf}. These material parameters can be obtained normally by experimental measurements or $\emph{ab initio}$ calculations~\cite{OSipr,LQLiu}. Due to technical difficulties, experimentally getting the temperature dependence of $A$ at the RE/TM interface among these four basic material parameters remains challenging.

Here, we report a simple electrical technique to derive temperature dependence of exchange stiffness $A$ in the artificially layered ferrimagnet consisting of a Py/Gd bilayer. First, the equilibrium orientation $\beta$ of the near uniformly magnetized Py sublayer can be obtained experimentally from angle-dependence of anisotropic magnetoresistance (AMR) curves measured at different temperatures and low magnetic fields much smaller than the effective exchange field $H_{ex}$. Second, using a modified Stoner-Wohlfarth model, the magnitude of the two relatively small exchange stiffness $A_{Py-Gd}$ and $A_{Gd}$ as the fitting parameters can be determined by directly comparing the experimentally obtained and theoretically calculated $\beta$ of the Py sublayer for four independent Py($t_{Py}$)/Gd(15 nm) bilayer films. This method enables the derivation of $A_{Py-Gd}$ and $A_{Gd}$ over a relatively wide temperature range until the temperature where the Gd-core sublayer begins to lose its long-range magnetic order. The obtained temperature dependences of $A_{Py-Gd}$ and $A_{Gd}$ exhibit a scaling power law with the magnetization of Gd with exponent $\kappa\approx 1.87$, close to the value for 3$d$ ferromagnetic metals with the generic simple cubic lattice and FePt alloy determined by domain wall and spin-wave approaches~\cite{Atxitia}. Finally, the experimental results can be well reproduced by micromagnetic simulations with the obtained parameters $A_{Py-Gd}$ and $A_{Gd}$, further confirming the reliability of this method.

\begin{figure*}[th]
  \includegraphics[width=1\textwidth]{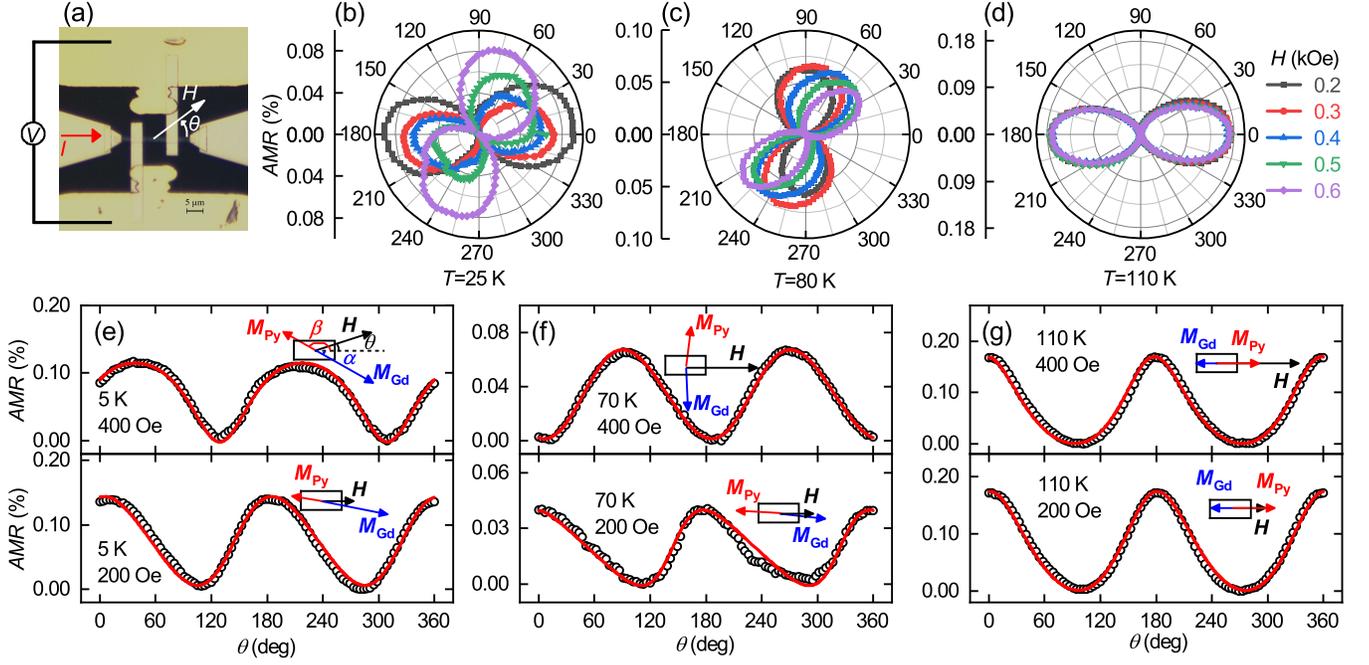}
  \caption{Representative dependence of resistance on magnitude and angle of the field for the Py(5)/Gd(15) bilayer. (a) Optical image of the devices and experimental setup of the angular-dependent AMR measurement. $\theta$ is defined as the angle between $H$ and $I$. (b)-(d) Polar plots of several representative angular dependent $R$ curves obtained with different labeled field values at $T$ = 25 K (b), 80 K (c) and 110 K (d). (e)-(g) AMR experimental data (black circles) and the corresponding fitting curves (solid red lines), as defined in the text. The insets illustrate the orientation of the applied in-plane magnetic field $H$ (black arrow), the magnetization of the Gd layer $M_{Gd}$ (blue arrow) and the Py layer $M_{Py}$ (red arrow). AMR ratio is defined by AMR = $\frac{R(\beta)-R_{90^o}}{R_{90^o}}$.}
  \label{fig1}
\end{figure*}

The Ni$_{81}$Fe$_{19}$ (Py)($t_{Py}$)/Gd(15 nm) bilayer films were deposited on an annealed sapphire substrate by dc magnetron sputtering at room temperature. The thickness $t_{Py}$ of the Py layer was 5, 10, 15, and 20 nm. The bilayers were protected from being oxidized by the MgO(2 nm)/Ta(2 nm) capping layer. The multilayer films were patterned into a stripe of 7 $\mu$m $\times$ 2 $\mu$m for transport measurements by combining electron beam lithography and sputtering. The standard four-probe technique was used to measure the isothermal field-dependent AMR with the in-plane field orientation at cryogenic temperatures[Fig.~\ref{fig1}(a)]. Figure~\ref{fig1} shows several representative angular dependent resistance curves of the Py(5)/Gd(15) bilayer obtained at different values of $H$ and $T$. At the Py-dominated high-temperature region $T > T_{comp}\simeq$ 100 K, all curves with $H$ far above its coercivity $H_c$ show the same sinusoidal dependence of $R$ on the orientation of $H$ with a period of 180$^o$ [Fig.~\ref{fig1}(d)], well consistent with the AMR of the single Py film, which indicates the direction of Py magnetization aligning with the field. In contrast, for the Gd-dominated low-temperature phase $T < T_{comp}$, the twofold symmetry axis of the AMR exhibits an obvious deflexion, and its orientation depends on the magnitude of field and temperature [Figs.~\ref{fig1}(b)-(c) and Figs.\ref{fig1}(e)-(f)]. These distinct angular-dependent $R$ curves are related to the field-driven twisted phase in the adjacent Gd sublayer at $T < T_{comp}$ because the Gd is expected to have one order of magnitude smaller exchange stiffness $A$ than the Py ~\cite{QChen,HNagura}. However, the Gd layer did not directly generate the AMR signal of the bilayer because it exhibits a negligible AMR ratio compared to the Py sublayer. Therefore, the Py sublayer can be used as an ideal angle sensor of the twisted state because Py has uniform magnetization in the field-driven twisted state, a very small $H_C$ ($\sim$ less than 10 Oe even at 5 K) and a sizeable AMR effect~\cite{liu2014}. From the change of $R(\theta)$ curves as a function of $H$, we can probe spin-flop transition and directly determine it critical field. The AMR experimental results of the Py/Gd bilayer, as shown in Figs.\ref{fig1}(e)-(g), can be described by a sinusoidal function of the magnetization orientation of the Py sublayer response to the direction of the current $I$ as follows~\cite{DChen}
\begin{equation}\label{AMR}
  R(\beta)=R_{90^o}+(R_{0^o}-R_{90^o})\cos^{2}(\beta)
\end{equation}
, where $\beta$ is the angle between \textbf{\emph{M}} of Py and the direction of $I$, $R_{90^o}$ and $R_{0^o}$ are the resistances at $\beta$ = 90$^o$ and 0, respectively. The equilibrium orientation of \textbf{\emph{M}} of Py in the Gd/Py bilayer is related to the values of the external magnetic field $H$, as well as the interlayer antiferromagnetic exchange stiffness A$_{Py-Gd}$ at Py-Gd interface.

\begin{figure*}[thb]
\centering
\includegraphics[width=1\textwidth]{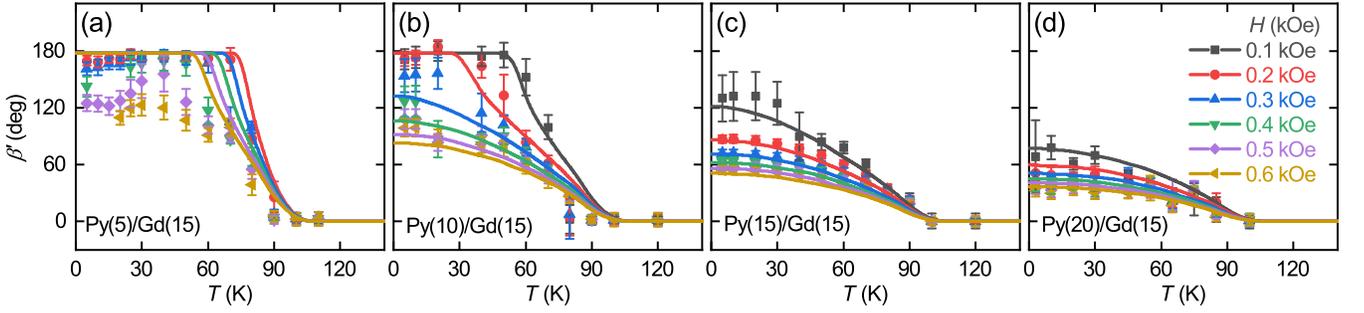}
\caption{Temperature-dependent angle $\beta^{\prime}$ between $\mathbf{M}_{Py}$ and $H$ under labeled field values for four different bilayer films Py(5)/Gd(15) (a), Py(10)/Gd(15) (b), Py(15)/Gd(15) (c), and Py(20)/Gd(15) (d). Symbols: $\beta^{\prime}=\beta-\theta$ is directly extracted from the experimental angular-dependence of resistance curves in Fig.~\ref{fig1} using the extended Stoner-Wohlfarth model. The error bars are determined by the standard deviations between the experimental R($\beta$) data and SW model fittings. The solid curves are the theoretically calculated results using the fitting parameters $A_{Gd}$ and $A_{Py-Gd}$. }\label{fig2}
\end{figure*}

We can extract the exchange stiffness inside the Gd-core sublayer and Py/Gd interface from the experimentally obtained AMR curves. Next, let us look more into how to do that. We first adopt the mean-field model to analyze the free energy of the Py/Gd bilayer by dividing the Py/Gd bilayer into four parts: the near-uniform magnetization Py layer $t_{Py}$, the intermediate interface region with an effective spin-spin correlation length $t_{Py-Gd}$, the twisted state region inside the Gd sublayer $L$ near the Py/Gd interface and the residue Gd layer $t_{Gd}-L$. Finally, the total free energy of the Py/Gd bilayer can be expressed as follows~\cite{Higgs}:
\begin{equation}
\begin{split}
&E_{total} = \frac{1}{2}\mu_{0}N_{yy}[M_{Py}^{2}\sin^{2}\beta t_{Py} + \int^{t_{Gd}}_{0}M_{Gd}^{2}\sin^{2}(\alpha(z))dz]\\*
&-\mu_{0}H[M_{Py}t_{Py}\cos(\beta^{\prime})+\int_{0}^{t_{Gd}}M_{Gd}\cos(\theta-\alpha(z))dz]\\*
&+ \frac{2A_{Py-Gd}t_{Py-Gd}}{a^{2}}\cos(\beta-\alpha_{0})+\frac{A_{Gd}}{L}(\alpha_{0}-\theta)^{2}
\label{StonerWohlfarth}
\end{split}
\end{equation}
where $\beta^{\prime}=\beta-\theta$ is the angle between the Py magnetization $\mathbf{M}$ and the field direction. The total free energy $E_{total}$ comprises the demagnetizing energy, the Zeeman energy of both the Py and Gd layers, and the exchange energy of the intermediate interface region $t_{Py-Gd}$ and the twisted state region inside the Gd sublayer. Zeeman energy (second term) is proportional to the thickness of the magnetic layers $t_{Py}$, $t_{Gd}$ and the external magnetic field $H$. The interfacial exchange energy per unit area at the interface of Py/Gd is independent of the thickness of the bilayer. The final term is the exchange energy of the twisted phase inside the Gd sublayer with depth $L$ near the Py-Gd interface. The interatomic distance $a\approx 0.35$ nm and the spin-spin correlation length $t_{Py-Gd}$ = 2 nm are used for the Py/Gd bilayer. $N_{yy}$ is the demagnetizing factor of the Py/Gd stripe. $\alpha$ ($\alpha_{0}$) is the angle between the direction of current and the magnetic moment in the Gd-core sublayer (at the intermediate interface between Py and Gd), as defined in the inset of Fig.~\ref{fig1}(e). $M_{Gd}$ and $M_{Py}$ are the saturation magnetization of the Gd-core sublayer and Py layers, determined separately by the SQUID magnetometer. Based on the previous resonant x-ray magnetic scattering experiment~\cite{Hosoito}, the interfacial Gd magnetization near the Py layer $M_{Gd\_int} \sim$ 6 $\mu_B$ is adopted for fitting. Finally, the equilibrium orientation of $\mathbf{M}_{Py}$ can be determined by minimizing the total free energy $E_{total}$ with respect to $\beta$, yielding $R(\beta)=R_{90^o}+(R_{0^o}-R_{90^o})\cos^{2}(\beta = \theta + f(\theta, H, A, ...))$, a result of a modified Stoner-Wohlfarth (SW) model~\cite{Stoner}. Figures~\ref{fig1}(e)-(g) show that the experimentally obtained AMR curves of the Py/Gd bilayer with several different $H$ at a specific temperature can be well fitted with this modified SW model with $L$, $\alpha_{0}$ and $A_{Gd}$ and $A_{Py-Gd}$ as the same free fitting parameters.

To further evaluate the reliability of the obtained $A_{Gd}$ and $A_{Py-Gd}$ values, we performed a direct comparison between the theoretical calculation and the experimentally obtained $\beta^{\prime}$ from AMR curves with different $H$ and $T$. Figures~\ref{fig2}(a)-(d) show temperature dependencies of experimentally obtained $\beta^{\prime}$ (symbols) and theoretically calculated results (solid lines) for all four studied Py($t_{Py}$ = 5, 10, 15, 20 nm)/Gd(15) bilayer samples with six different $H$ = 100, 200, 300, 400, 500 and 600 Oe at $\theta$ = 0. Although $\beta$ has some deviation between the experiment and calculation for Py(5)/Gd(15) sample at low-temperature range, it still exhibits a good consistency with the theoretical results if one considers the overall results of all four samples with different Py thicknesses and six different fields using the same $A_{Gd}$ and $A_{Py-Gd}$. The deviation for Py(5)/Gd(15) is likely related to non-negligible shape magnetic anisotropy and pinning effects due to some inhomogeneities and defects that existed in the thinner Py sublayer. All samples show that $\mathbf{{M}_{Py}}$ perfectly parallels $H$ ($\beta^{\prime}$ = 0) at $T$ $>$ 110 K, suggesting that the Py layer dominates the magnetization of the bilayers, consistent with a low Curie temperature $T_C$ = 130 $\pm$ 10 K for the 15 nm thick Gd sublayer, determined by its M-T curve (see supplementary material Fig.S1). However, a short-range exchange would exist in the Gd-core layer, and a large interface magnetization $M_{Gd\_int}$ may be induced by the magnetic proximity effect(MPE) above this temperature~\cite{Hosoito}. The angle $\beta^{\prime}$ between $\mathbf{M}_{Py}$ and $H$ decreases with increasing the Py thickness, and as well as $H$, which is attributed to the increase of the Zeeman energy $\mu_{0}\mathbf{M}_{Py}\cdot H t_{Py}$ and almost unchanged exchange energy in the Py/Gd interface because the studied four Pt/Gd samples have the same 15 nm thick Gd sublayer.

\begin{figure}[h]
\centering
\includegraphics[width=0.47\textwidth]{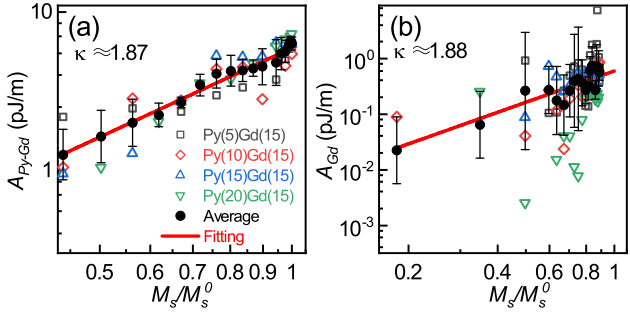}
\caption{The logarithmic plots of the interlayer exchange stiffness $A_{Py-Gd}$ between Gd and Py (a) and bulk exchange stiffness $A_{Gd}$ inside Gd layer as a function of the normalized magnetization $M_{s}/M^{0}_{s}$ of the Gd layer by the saturation magnetization $M^{0}_{s}$ at $T$ = 4 K. The hollow symbols are the obtained experimental results for the studied samples with four different Py thicknesses, the solid circles represent their average value and the solid red line is a linear fitting with the scaling exponent in the main text. }\label{fig3}
\end{figure}

Based on the above-discussed analysis of the angular dependence of AMR curves obtained at different $T$, we can quantitatively calculate the temperature dependence of the interfacial exchange stiffness $A_{Py-Gd}$ between Py and Gd and the bulk $A_{Gd}$ inside the Gd thin film. We plot all obtained $A_{Gd}$ and $A_{Py-Gd}$ for the four studied samples at different $T$ as a function of the normalized magnetization $M_s/M_s^0$ of the Gd layer, as shown in Figs.~\ref{fig3}(a) and ~\ref{fig3}(b), respectively. Figure~\ref{fig3}(a) shows that the average value of $A_{Py-Gd}$ vs. $M_{s}/M^{0}_{s}$ can be well fitted by the scaling exponent $A = A_{0}(M_{s}/M^{0}_{s})^{\kappa}$ with exponent $\kappa\approx1.87$, close to previously reported $\kappa= 1.66$ for the generic simple cubic lattice and $\kappa= 1.76$ for a FePt through domain wall and spin-wave approaches~\cite{Atxitia}. Note that the saturation magnetization $M_{Py}$ can be considered as a constant for our studied temperature range due to a very high Curie temperature ($T_{C} >$ 680 K) for the Py sublayer~\cite{Mauri,CMFu}. In addition, the obtained average value of $A_{Gd}$ as a function of $M_{s}/M^{0}_{s}$ also follows this scaling function with the same exponent but with a considerable error uncertainty. The reason is that the value of $A_{Py-Gd}$ is one order of magnitude larger than that of $A_{Gd}$.
Therefore, $A_{Py-Gd}$ is the dominant weight factor in determining Stoner-Wohlfarth formula obtained by minimizing the total free energy $E_{total}$ with respect to $\beta$.
The obtained large derivation of $A_{Gd}$ is reasonable because a tiny fitting deviation in $A_{Py-Gd}$ would cause a substantial error uncertainty in $A_{Gd}$.

\begin{figure}[hb]
\includegraphics[width=0.47\textwidth]{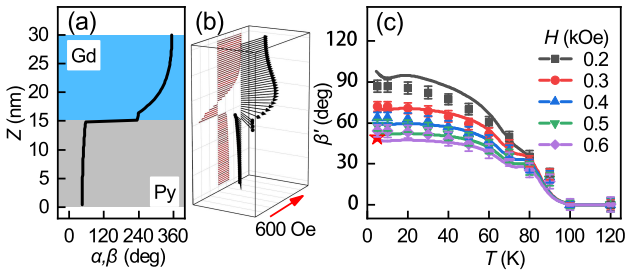}
\caption{Micromagnetic simulated results for the Py(15)/Gd(15) bilayer. (a) - (b) Two- (a) and three-dimensional (b) schematics of depth-dependent magnetization M$_{Py}$ and M$_{Gd}$ angle profiles under an external magnetic field $H$ = 0.6 kOe and $T$ = 5 K, marked as the red star in (c). (c) Comparison between experimentally obtained (symbols, the same date in Fig.~\ref{fig2}(c)) and simulated (solid curves) temperature-dependence of angle $\beta^{\prime}$ with different labeled fields. The error bars are defined by the standard deviations between the experimental $R(\beta)$ data in Fig.~\ref{fig1} and SW model fittings.}\label{fig4}
\end{figure}

Although a well-fitting of temperature- and field-dependent AMR data and the expected scaling power-law with a reasonable exponent have been obtained in the discussion above, it is still necessary to check the validity of our method. We simulated the static magnetization configuration with atomic depth profiling of the ferrimagnetic bilayer thin film using Py(15)/Gd(15) as an example by using the OOMMF code~\cite{oommf}. In principle, the micromagnetic model can be simplified to a one-dimensional spin chain directed along with the thickness of the film. Thus, to reduce the simulation time, we divided a Py(15)/Gd(15) stripe of 6 $\mu$m $\times$ 1.8 $\mu$m into 2 $\mu$m $\times$ 0.6 $\mu$m $\times$ 0.4 nm cells to perform simulation. The micromagnetic parameters used in our simulations are the same as the fitting parameters obtained above, e.g., $M_{Py}$ = 860 emu/cm$^3$, $A_{Py}$ =10 pJ/m, $M_{Gd}$ = 1345 emu/cm$^3$, $M_{Gd\_int}$ = 1800 emu/cm$^3$ at 2 K, $M_{Gd}(T)$, $A_{Gd-Py}(T)$ and $A_{Gd}(T)$, which were shown in supplementary material Fig.S1 and Fig.~\ref{fig3}.

Figures~\ref{fig4}(a) and~\ref{fig4}(b) show that all spins form a near uniformed ferromagnetic state with a canted angle $\beta^{\prime}$ of 50$^o$ from the direction of $H$ at $H$ = 0.6 kOe and $T$ = 5 K. For the interface region within spin-spin correlation length $t_{Py-Gd}$ = 2 nm, the Gd layer also keeps an almost collinear spin state, and its spins are antiparallel to the adjacent Py sublayer $\mathbf{M}_{Py}$ ($\alpha_{Gd}\sim$ 230$^o$). In the residual Gd-core region, the external field can induce a twisted state with an angle canted by 130$^o$ to parallel $H$. It is easy to understand from the view of the discussed minimization of the total free energy $E_{total}$ under $H$ above. The field-driven twisted state inside the Gd-core sublayer can facilitate the further reduction of $E_{total}$ because $A_{Gd}$(5 K) = 0.6 pJ/m is almost one order of magnitude smaller than $A_{Py-Gd}$(5 K) = 5.9 pJ/m and $A_{Py}$ = 10 pJ/m. Furthermore, temperature- and field-dependence of the canted angle $\beta^{\prime}$ also have been studied systematically in our simulations, as shown (solid lines) in Fig.~\ref{fig4}(c). One can see that the simulation results agree well with the experimental results (represented by symbols), especially for the large $H$.

In summary, we demonstrate that the abundant field-driven spin-flop transition, canted and twisted magnetization states in the artificially layered ferrimagnet consisting of a Py/Gd bilayer can be well characterized by performing temperature and angle of magnetic field dependencies of AMR using the Py sublayer as detection sensor. We quantitatively estimate the detailed temperature dependence of exchange stiffness $A_{Py-Gd}$ and $A_{Gd}$ from those experimentally obtained angle dependence of AMR at different temperatures and fields in combination with a modified Stoner-Wohlfarth model. The obtained low-temperature exchange stiffness $A_{Py-Gd}$ and $A_{Gd}$ have a scaling power low with the Gd magnetization with exponent $\kappa\approx1.87$. We also have reproduced the experimental results by micromagnetic simulations with the experimentally obtained parameters $A_{Py-Gd}$ and $A_{Gd}$. Our demonstrated AMR method can easily access various experimental conditions, e.g., cryogenic temperature, magnetic fields, and microscale and/or nanoscale samples. It enables us to electrically detect the spin-flop transition, canted and twisted magnetization states and their dynamics in the artificially interlayered ferrimagnet-based nanodevices consisting of 3\emph{d} and 4\emph{f} FM metals.

See the supplementary material for additional details on the magnetic susceptibility characteristics of the Py(5)/Gd(15) bilayer and the single Gd and Py layers.

We acknowledge support from the National Natural Science Foundation of China (No. 12074178, No. 12004171), the Applied Basic Research Programs of Science and Technology Commission Foundation of Jiangsu Province (Grant No. BK20200309), the Open Research Fund of Jiangsu Provincial Key Laboratory for Nanotechnology, and the Scientific Foundation of Nanjing University of Posts and Telecommunications (NUPTSF) (Grant No. NY220164).

$\mathbf{DATA AVAILABILITY}$
The data that support the findings of this study are available from the corresponding author upon reasonable request.

\bibliography{manuscriptREF}
\bibliographystyle{apsrev4-1}

\end{document}